\documentclass[11pt,a4paper]{article} 
\usepackage{jcappub}
\usepackage{lineno}

\title{Effects of uncertainties in simulations of extragalactic UHECR propagation, using CRPropa and {\it SimProp} }

\author[a]{R. Alves Batista,}
\author[b]{D. Boncioli,}
\author[c]{A. di Matteo,}
\author[a,d]{A. van Vliet}
\author[e]{and D. Walz}

\affiliation[a]{II. Institut f\"ur Theoretische Physik, Universit\"{a}t Hamburg,\\
Hamburg, Germany}
\affiliation[b]{INFN, Laboratori Nazionali del Gran Sasso,\\
Assergi (L'Aquila), Italy}
\affiliation[c]{Dipartimento di Scienze Fisiche e Chimiche dell'Universit\`{a} dell'Aquila and INFN,\\
L'Aquila, Italy}
\affiliation[d]{Department of Astrophysics, IMAPP, Radboud University Nijmegen,\\
Nijmegen, the Netherlands}
\affiliation[e]{RWTH Aachen University, III. Physikalisches Institut A,\\
Aachen, Germany}

\emailAdd{rafael.alves.batista@desy.de}
\emailAdd{denise.boncioli@lngs.infn.it}
\emailAdd{armando.dimatteo@aquila.infn.it}
\emailAdd{a.vanvliet@astro.ru.nl}
\emailAdd{walz@physik.rwth-aachen.de}

\abstract{The results of simulations of extragalactic propagation of ultra-high energy cosmic rays (UHECRs) have intrinsic uncertainties due to poorly known physical quantities and approximations used in the codes. We quantify the uncertainties in the simulated UHECR spectrum and composition due to different models of extragalactic background light (EBL), different photodisintegration setups, approximations concerning photopion production and the use of different simulation codes. We discuss the results for several representative source scenarios with proton, nitrogen or iron at injection. For this purpose we used {\it SimProp} and CRPropa, two publicly available codes for Monte Carlo simulations of UHECR propagation. CRPropa is a detailed and extensive simulation code, while {\it SimProp} aims to achieve acceptable results using a simpler code. We show that especially the choices for the EBL model and the photodisintegration setup can have a considerable impact on the simulated UHECR spectrum and composition.
}


\begin{document}
\maketitle
\flushbottom
\section{Introduction \label{sec.intro}}
The most energetic particles in the universe, the ultra-high energy cosmic rays (UHECRs), have been the subject of intense research for over fifty years. Although considerable progress in this field has been made in recent years with results from the two largest cosmic ray experiments, the Pierre Auger Observatory and the Telescope Array, the origin of these particles remains a mystery. The only aspect that has been widely agreed on is that the vast majority of the most energetic cosmic rays are charged particles (protons and/or other atomic nuclei) originating from outside our galaxy, and therefore they can interact with intergalactic background radiation and/or magnetic fields during their propagation to Earth, hence affecting their properties.

At energies above $10^{18}\,$ eV the cosmic ray spectrum presents some interesting features. One of these is the so-called ``ankle'', at $E \approx 5 \times 10^{18}\,$eV, which is a flattening of the measured spectrum. Another feature is the flux suppression above $\approx 2 \times 10^{19}\,$eV~\cite{Abraham:2008ru,Abbasi:2007sv}. The flux suppression may be a consequence of the well-known Greisen-Zatsepin-Kuzmin (GZK) effect~\cite{Greisen:1966jv,Zatsepin:1966jv}, due to the interaction of UHECRs with the cosmic microwave background~(CMB). Pure proton models implementing the GZK effect, where the ankle is explained by proton interaction signatures as well, have been proposed extensively (see e.g. refs.~\cite{Hill:1983mk,Berezinsky:2002nc,Aloisio:2006wv}). However, the flux suppression could also be a signature of photodisintegration of nuclei by the CMB and the extragalactic background light (EBL), or of the maximum acceleration energy attainable by UHECR sources, as suggested in the context of the ``disappointing model''~\cite{Aloisio:2009sj}. The need for a vanishing proton component at $E \gtrsim 10^{19}$~eV is driven by the increasingly heavy and pure mass composition as measured by the Pierre Auger Observatory~\cite{Aab:2014aea}. In recent years combined investigations of the spectrum and composition measurements of Auger have shed a new light on the subject (see e.g. refs.~\cite{Taylor:2011ta,Aloisio:2013hya,Fang:2013cba,Taylor:2013gga,Peixoto:2015ava,Globus:2015xga,Taylor:2015rla,Unger:2015laa}).

Possible models of
UHECR sources can have many unknown parameters, such as their
distribution, the maximum acceleration energy, the shape of the
injection spectrum and the initial mass composition of cosmic rays.
Therefore, to be able to thoroughly study UHECR source models and
ascertain their compatibility with the available data, it is necessary
to simulate the propagation of UHECRs in scenarios covering a large parameter space. To do this efficiently fast computational tools are required. There are a number of public codes for the propagation of UHECRs, including CRPropa~\cite{Armengaud:2006fx,Kampert:2012fi,crpropa3}, {\it SimProp}~\cite{Aloisio:2012wj}, TransportCR~\cite{Kalashev:2014xna} and HERMES~\cite{DeDomenico:2013psa}, as well as private codes (see e.g. refs.~\cite{Allard:2005ha,Hooper:2006tn,Aloisio:2010he}), available for that purpose.

The simulated UHECR spectra and mass compositions might depend strongly on poorly known quantities such as the spectrum and evolution of the EBL and photodisintegration cross sections, as well as on different computational treatments and approximations made in the different simulation codes. The main goal of the present work is to investigate how sensitive the simulations are to different models of the EBL, to different ways of treating photopion production, and to different photodisintegration implementations, as well as to compare the CRPropa and {\it SimProp} codes, aiming to understand how different computational treatments and approximations can affect the outcome of simulations.

This paper is structured as follows: in section~\ref{sec.propagation} the propagation of UHECRs in the universe is discussed; in section~\ref{sec.MCcodes} the simulation codes are described; in section~\ref{sec.comparisons} the results of the comparisons are presented, for the cases of proton propagation and nuclei propagation separately; finally, in section~\ref{sec.discussion} the results are discussed and the conclusions are presented.

\section{UHE cosmic ray propagation \label{sec.propagation}}
When propagating in the extragalactic space, UHECRs interact with photon backgrounds, namely the CMB and the EBL. The CMB has a blackbody spectrum with temperature at redshift $z$ given by $T(z) = T_0 (1+z)$, with $T_0=2.725\,$K. The EBL encompasses electromagnetic radiation in a wide range of frequencies, from infrared to ultraviolet, and is poorly known mainly due to uncertainties concerning its time evolution\footnote{For details concerning the EBL models used in this work, refer to appendix~\ref{sec.EBLmodels}.}.

The mean free path~$\lambda$ for a particle with Lorentz factor $\Gamma$ interacting at redshift~$z$ with a diffuse photon background of spectral number density $n(\epsilon,z)$ for photons with energy~$\epsilon$ (in the laboratory frame) can be written as
\begin{equation}
	\lambda^{-1}(\Gamma,z) = \frac{1}{2\Gamma^2} \int \limits_{0}^{\infty} \int \limits^{2\Gamma \epsilon}_0 n(\epsilon,z) \frac{1}{\epsilon^2} \epsilon' \sigma(\epsilon') \,d\epsilon' d\epsilon, \label{eq:lambda}
\end{equation}
where $\sigma(\epsilon^\prime)$ is the cross section for a given interaction between the cosmic ray and photons of energy $\epsilon^\prime = (1-\cos\theta) \Gamma \epsilon$ (in the nucleus' rest frame), with $\theta$ being the angle between the photon and the particle's momentum in the laboratory frame. In the energy range of interest ($E \gtrsim 10^{18}\,$eV), the most important of such processes are photopion production, pair production and photodisintegration.

Photopion production is the process in which a nucleon (free or bound to a nucleus) interacts with a background photon producing nucleons and pions ($p+\gamma \rightarrow p+\pi^0$, $p+\gamma \rightarrow n + \pi^+$, etc). Charged pions produce neutrinos and electrons ($\pi^+ \rightarrow \mu^+ + \nu_\mu$, $\mu^+ \rightarrow e^+ + \nu_e + \bar\nu_\mu$), whereas neutral pions produce gamma rays ($\pi^0 \rightarrow \gamma + \gamma$). The threshold energy for photopion production is $\sim 3 \times 10^{19} (\mathrm{meV}/ \epsilon)\,$eV, where $\epsilon$ is the energy of the photon. The dominant photon background at ultra-high energies is the CMB ($\epsilon \approx 0.7\,$meV), causing the so-called Greisen-Zatsepin-Kuzmin (GZK) effect, which induces a cutoff in the UHECR spectrum starting from energies around $3\times10^{19}\,$eV. The GZK effect implies that almost all protons reaching Earth with energy greater than $10^{20}\,$eV must originate from within about 100~Mpc. For nuclei with mass number $A$, the threshold for photopion production is $A$ times that for protons, but photodisintegration (see below) is possible at lower energies.

Pair production is the process whereby electron-positron pairs are created due to the interaction of UHE nuclei with background photons ($^{A}_{Z}X + \gamma \rightarrow {}^{A}_{Z}X + e^+ + e^-$). It has a relatively short mean free path, but a very small fractional energy loss, thus being well approximated as a continuous energy loss (CEL) process. The threshold energy is $\sim 5\times 10^{17} A (\mathrm{meV}/\epsilon)~\mathrm{eV}$, where $A$ is the atomic mass of the nucleus. In the case of nuclei, the energy loss length ($\ell \equiv \Gamma / (d\Gamma / dx) $) scales as $\ell^{-1}_\text{nuclei} = \ell^{-1}_\text{protons} Z^2 / A$.

Photodisintegration occurs when a nucleus is split into smaller parts due to interactions with photons ($^A_Z X + \gamma \rightarrow {}^{A-1}_{Z}X + n$, $^{A}_{Z}X + \gamma \rightarrow {}^{A-1}_{Z-1}X + p$, etc). Cross sections for this interaction are dominated by the giant dipole resonance (GDR) for photons with energies $\epsilon^\prime \lesssim 30\,$MeV (in the nucleus rest frame). For $30\,\mathrm{MeV} < \epsilon^\prime < 150\,\mathrm{MeV}$ quasi-deuteron (QD) processes occur, typically causing the emission of multiple nucleons. For $\epsilon^\prime > 150\,$MeV photodisintegration cross sections rapidly vanish and pion production takes over. Daughter-nuclei approximately conserve the same Lorentz factors from their corresponding progenitor nuclei as nuclear recoil is negligible, for the energy of interest is much less than the rest mass of the nucleus.

Due to the expansion of the universe, all particles undergo adiabatic losses. In this case, the energy loss rate is given by
\begin{equation}
	-\frac{1}{\Gamma} \frac{d\Gamma}{dx} = \frac{1}{c}H(z) = \frac{H_0}{c}\sqrt{\Omega_{\rm m} + (1+z)^3 \Omega_\Lambda },
\end{equation}
where $H_0 \equiv H(0) \approx 70\,$km/s/Mpc is the Hubble constant at present time, $\Omega_{\rm m} \approx 0.3$ is the density of matter (baryonic and dark matter), and $\Omega_\Lambda \approx 0.7$ is the dark energy density, in the standard cosmological model ($\Lambda$CDM).

Another relevant process is nuclear decay. Unstable products of photodisintegration and photopion interactions can have lifetimes shorter than the typical propagation lengths involved, causing their decay along their trajectory to Earth. The most relevant nuclear decay processes for this energy range are $\alpha$ and $\beta^\pm$ decays, as well as nuclear dripping.

\begin{figure}[t]
	\includegraphics[width=0.495\columnwidth]{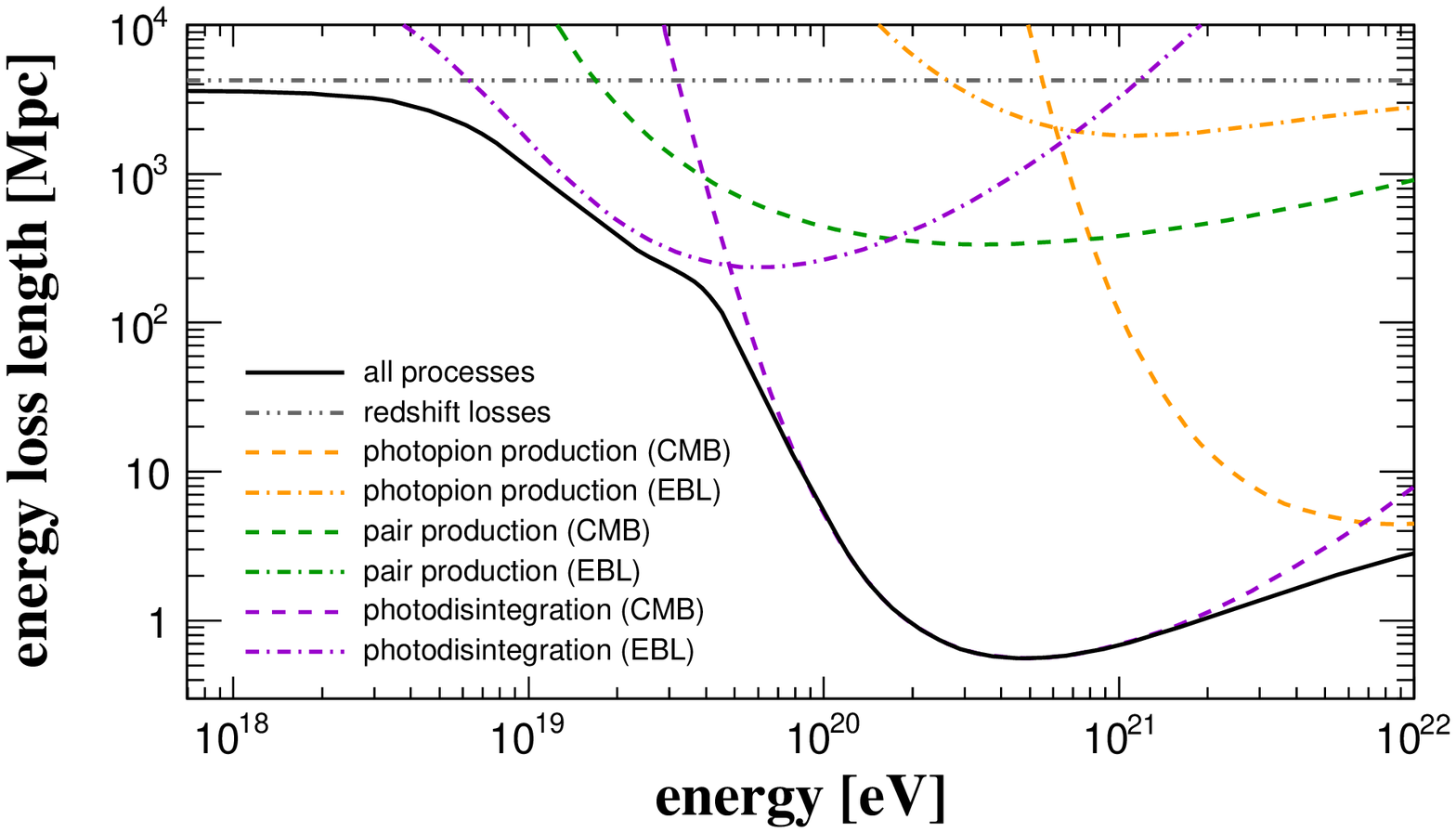}
	\includegraphics[width=0.495\columnwidth]{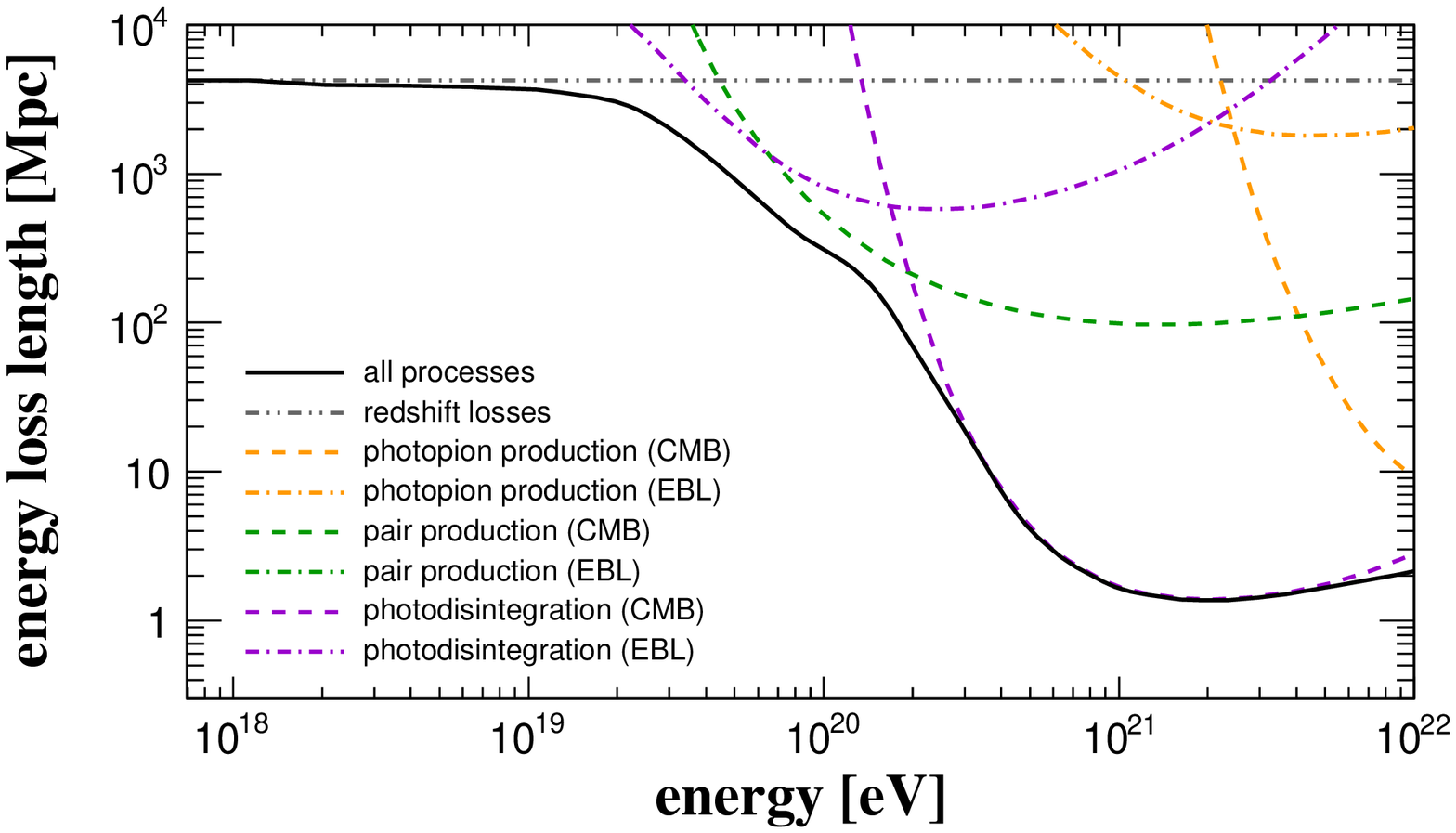}
	\caption{Energy loss length for $^{14}$N (left panel) and $^{56}$Fe (right panel) at $z=0$. Dotted-dashed lines correspond to interactions with the EBL, and dashed lines with the CMB. Green curves represent electron pair production, purple photodisintegration, and orange photopion production. Double dotted-dashed lines are the energy loss length for adiabatic losses (Hubble radius). The thick black curve is the combined energy loss length, taking into account all processes. This plot was obtained using the interaction rate tables from CRPropa 3, with the EBL model of Gilmore {\it et al.}~\cite{Gilmore:2011ks}.}
	\label{fig:ell}
\end{figure}

A compilation of the energy loss length for different processes for $^{14}$N and $^{56}$Fe can be seen in figure~\ref{fig:ell}.

Magnetic fields may affect cosmic ray observables measured on Earth, such as the spectrum and mass composition. These effects may be due to the presence of intervening intergalactic magnetic fields, the magnetization of the environment where the observer lies, the magnetization of sources and their surroundings, or any combination thereof. There are large uncertainties on the strength and distribution of magnetic fields, thus making it nontrivial to properly assess their impact on the UHECR spectrum and composition. On the other hand, in the case of sources uniformly distributed separated by a distance much smaller than the typical lengths involved (e.g. Larmor radius), it has been proven that the spectrum has a universal shape regardless of the properties of magnetic fields~\cite{Aloisio:2004jda}. A uniform source distribution is a reasonable assumption because the sources of UHECRs are not known, and considering the lower bounds on the density of sources estimated by the Pierre Auger Collaboration~\cite{Abreu:2013kif}. In refs.~\cite{Lemoine:2004uw,Sigl:2007ea,Globus:2007bi,Kotera:2007ca,Mollerach:2013dza,Batista:2014xza} it was claimed that spectrum and composition may be affected by magnetic fields, although the rigidity at which this is relevant is not clear, ranging from $10^{15}\,$V up to $\sim 10^{18}\,$V. For $E/Z \gtrsim 10^{18}\,$V most of the aforementioned works predict small or negligible effects due to magnetic fields.

\section{Monte Carlo codes\label{sec.MCcodes}}
In the following the CRPropa and {\it SimProp} codes are briefly described.
Attention is given to the considered models of the EBL and photodisintegration, as well as their implementation in the codes.

\subsection{The CRPropa propagation code}\label{sec:CRP}
CRPropa 3~\cite{crpropa3} (see also refs.~\cite{Batista:2013gka,Batista:2014vva,Batista:2014wya}) is a public\footnote{Code available at \url{http://crpropa.desy.de}.} Monte Carlo code for propagating UHE nuclei, gamma rays and neutrinos in the universe. It includes all relevant interactions in the energy range of $\sim 6 \times 10^{16}$ up to $10^{22}\,\mathrm{eV}$, as well as many magnetic field environments and source distribution configurations. Three propagation modes are available, namely one-dimensional (1D), three-dimensional (3D), and four-dimensional (4D) modes.
For the purposes of this work we will focus on the 1D mode in order to compare with the {\it SimProp} code, which is limited to 1D simulations.

The energy loss processes and interactions implemented in this case are pair production, photopion production, photodisintegration, nuclear decay and adiabatic losses. Pair production and adiabatic losses are treated as CEL processes, whereas photopion production, photodisintegration and nuclear decay are handled stochastically. The treatment of pair production follows ref.~\cite{Blumenthal:1970nn}, and photopion production is handled by the SOPHIA code~\cite{Mucke:1998mk}.

In CRPropa 3 photonuclear cross sections are obtained from the TALYS 1.6 code~\cite{talys} using the parameters described in appendix~\ref{sec.GDRparams}, in contrast to CRPropa 2~\cite{Kampert:2012fi}, which uses TALYS 1.0 with default parameters. TALYS has been used to predict photodisintegration products of all available exclusive channels: proton, neutron, deuterium, tritium, helium-3 and helium-4 (alpha particle) and combinations thereof. Nuclei with $A<12$ are not treated with TALYS. Instead, their photonuclear cross sections are compiled from various references. For $^2$H, $^3$H, $^3$He, $^4$He, $^9$Be they are obtained from ref.~\cite{rachen1996}. In the case of $^3$H and $^3$He they are scaled by factors 1.7 and 0.66, respectively, with respect to ref.~\cite{rachen1996}. The parametrizations for $^9$Be is refitted to data, as shown in ref.~\cite{nierstenhoefer2011}. Cross sections for $^{6}$Li, $^{8}$Li, $^{7}$Be, $^{11}$Be, $^{8}$B, $^{10}$B, $^{11}$B, $^{10}$C and $^{11}$C are taken from ref.~\cite{kossov2002}. Cross sections for $^7$Li are obtained by interpolation of experimental data~\cite{kulchitskii1963,varlamov1986}. For these nuclei, one proton is lost if $Z > N$, where $N$ is the number of neutrons, one neutron if $N > Z$, or one of them with equal probability if $N=Z$. In CRPropa, a total of 185 isotopes ($Z \leq 26$, $N \leq 30$ with a lifetime $\tau > 2\,$s) and 2220 disintegration channels are considered. Alternatively to the TALYS model, the total cross sections from ref.~\cite{kossov2002} can be used, while keeping the branching ratios from TALYS.

Several EBL models are implemented in CRPropa~3, namely the ones by Kneiske {\it et al.}~\cite{Kneiske:2003tx}, Stecker {\it et al.}~\cite{Stecker:2005qs,Stecker:2006eh}, Franceschini {\it et al.}~\cite{Franceschini:2008tp}, Finke {\it et al.}~\cite{Finke:2009xi}, Dom{\'i}nguez {\it et al.}~\cite{Dominguez:2010bv} and Gilmore {\it et al.}~\cite{Gilmore:2011ks}.
The implementation of photopion production, photodisintegration and electron pair production is based on tabulated mean free path data calculated with the comoving photon density $n(\epsilon,z=0)$ at redshift $z=0$.
By approximating a redshift independent spectral shape of the photon density, the following scaling relation for the mean free path $\lambda(\Gamma,z)$ at redshift $z$ can be used (cf. \cite{Kampert:2012fi})
\begin{align}
\lambda(z) = (1+z)^3 \frac
        {\int_0^\infty n(\epsilon,z)d\epsilon}
        {\int_0^\infty n(\epsilon,0)d\epsilon}
        \lambda((1+z)\Gamma,z=0)
\end{align}
For photopion production CRPropa~3 alternatively provides redshift-tabulated values of the mean free path $\lambda(\Gamma,z)$ instead of the scaling relation.

The treatment of nuclear decays is based on the NuDat 2.6 database\footnote{For details refer to the website \url{http://www.nndc.bnl.gov/nudat2/}.}, which provides data for decay channels and nuclear lifetimes.
In $\beta^+$ decays the absence of electron capture for fully ionized cosmic ray nuclei is accounted for in CRPropa.

\subsection{The {\it SimProp} propagation code}

The original version of {\it SimProp},\footnote{{\it SimProp} is available upon request to \href{mailto:SimProp-dev@aquila.infn.it}{\nolinkurl{SimProp-dev@aquila.infn.it}}.} described in ref.~\cite{Aloisio:2012wj,boncioli2011} and henceforth referred to as {\it SimProp}~v2r0, can simulate the one-dimensional propagation of protons and nuclei in the absence of magnetic fields. All particles undergo adiabatic energy losses and pair production losses, treated as CEL with rates computed as in ref.~\cite{Berezinsky:2002nc}. Protons can lose energy via photopion production, approximated as a CEL with rates computed as in ref.~\cite{Berezinsky:2002nc}; pion production is neglected altogether in the propagation of nuclei. Nuclei undergo photodisintegration, which is treated stochastically according to the Puget-Stecker-Bredekamp (PSB) model~\cite{Puget:1976nz} with Stecker-Salamon energy thresholds~\cite{Stecker:1998ib} (see also appendix~\ref{sec.photodis}); all the ejected nucleons are treated as protons and the residual nucleus is treated as the stable isobar for the corresponding value of $A$. This version of the code computes interactions on the Stecker et al. EBL model~\cite{Stecker:2005qs,Stecker:2006eh} (see also appendix~\ref{sec.EBLmodels}) as interpolated on a 2D grid of photon energies and redshifts, or a power-law approximation thereof~\cite{Aloisio:2008pp,Aloisio:2010he}

An updated version, {\it SimProp}~v2r1~\cite{Aloisio:2013kea}, treats pion production on the CMB as a stochastic process, both for protons and for nuclei, with rates computed using total cross sections from SOPHIA~\cite{Mucke:1998mk}; for nuclei, the rate is approximated as $A$ times that for a proton with the same Lorentz factor; all photohadronic processes are treated as single-pion production, with branching ratio $1/3$ for neutral pions and $2/3$ for charged pions, in accord with isospin invariance. Also, the type of the nucleons ejected in photodisintegration is now randomly chosen and the residual nucleus is selected according to conservation of electric charge, so that photodisintegration and photopion production can now produce neutrons and unstable nuclei, though these are assumed to immediately undergo beta decay. Neutrinos produced in the decay of pions, neutrons and unstable nuclei are also tracked.

{\it SimProp}~v2r2~\cite{Aloisio:2015sga} also optionally considers pion production on the EBL, whose effect on UHECR fluxes is very small but relevant for cosmogenic neutrino production at energies below $\sim 1$ EeV.  It also allows the user to choose the Kneiske et al. EBL model~\cite{Kneiske:2003tx} using the photon energy at $z=0$ scaled by a $z$-dependent factor as in CRPropa, as well as the ones used in {\it SimProp}~v2r1. Moreover, it fixes a bug affecting the low energy tail of neutrino spectra, and uses a faster implementation to calculate interaction lengths resulting in much shorter computation times.

In this work we use an extended version of {\it SimProp} (provisionally named {\it SimProp}~v2r3) which will be released in the near future. {\it SimProp}~v2r3 includes several new models of EBL, including Dom\'inguez et al.~\cite{Dominguez:2010bv} and Gilmore et al.~\cite{Gilmore:2011ks} (see also appendix~\ref{sec.EBLmodels}), all interpolated on a 2D grid. Furthermore, it allows the user to specify different models of photodisintegration. In this work we use photodisintegration cross sections obtained from TALYS 1.6 with the settings described in appendix~\ref{sec.GDRparams}, i.e. the same settings as used for CRPropa~3.

    \label{sec:v2r3}
    We used the following approximation scheme to implement TALYS cross sections in {\it SimProp}~v2r3, because while the uncertainties associated with the $\ln A$ measurements~\cite{Aab:2014kda} by the Auger observatory
    are in principle small enough to distinguish protons from helium-4, they
    are too large to distinguish consecutive intermediate nuclei, e.g. carbon-12
    from carbon-13. Therefore it is important that UHECR propagation simulations
    accurately predict the number of protons and $\alpha$-particles reaching Earth,
    but it is unnecessary to have the correct distribution of individual intermediate masses.
    In {\it SimProp}~v2r3, therefore, only two photodisintegration processes are
    implemented: nucleon ejection and $\alpha$ particle ejection; the interaction rates
    for these processes are taken to be the sum of all actual processes weighted
    by the number of nucleons and $\alpha$-particles ejected, respectively. This
    ensures that the numbers of free nucleons and of $\alpha$-particles at Earth, assuming
    that the interaction rates don't change too rapidly with $z$ or $A$, are reproduced in good approximation\footnote{
        For example, assume we have $^{14}$N nuclei originating 70 Mpc away, and the
        only relevant process is $^{14}\mathrm{N} + \gamma \to ^{12}\mathrm{C} + \mathrm{p} + \mathrm{n}$
        with interaction length 100 Mpc. A fraction $\exp(-0.7) \approx 50\%$ of the nuclei will survive,
        and at Earth, for each 100 $^{14}\mathrm{N}$ nuclei injected, we will have
        in average 50 $^{14}\mathrm{N}$ nuclei, 50 $^{12}\mathrm{C}$ nuclei, and
        100 protons. If we chose to approximate this process as $^{14}\mathrm{N} + \gamma \to ^{13}\mathrm{C} + \mathrm{p}$
        and $^{13}\mathrm{C} + \gamma \to ^{12}\mathrm{C} + \mathrm{n}$ with interaction
        length 50 Mpc each, a fraction $\exp(-0.7)^2 \approx 25\%$ of the nuclei will survive,
        $2\exp(-0.7)(1-\exp(-0.7)) \approx 50\%$ will interact once, and $(1-\exp(-0.7))^2 \approx 25\%$ will interact twice,
        and at Earth, for each 100 $^{14}\mathrm{N}$ nuclei injected, we will have
        in average 25 $^{14}\mathrm{N}$ nuclei, 50 $^{13}\mathrm{C}$ nuclei, 25 $^{12}\mathrm{C}$ nuclei, and
        100 protons. Both the number of protons and the average mass of the intermediate nuclei will then be
        well approximated, though the numbers of individual intermediate nuclides will be different.
    }.
    Since deuterium, tritium and helium-3 have very short disintegration rates,
    such ejectiles are treated as collections of free nucleons; also, since neutrons
    have a short decay length except at extremely high energy and even then the
    air showers they produce are indistinguishable from those of protons, we treat
    all nucleons as the same.
    Concretely, {\it SimProp}~v2r3 considers these two processes:
    \begin{itemize}
        \item nucleon ejection, with cross section
        $$\sigma_N = \sum_\text{channels}
        n_N \sigma_{n_\mathrm{n}n_\mathrm{p}n_\mathrm{d}n_\mathrm{t}n_\mathrm{h}n_\alpha} = \langle n_N \rangle \sigma_\text{tot},$$
        where $n_N = n_\mathrm{n} + n_\mathrm{p} + 2n_\mathrm{d} + 3n_\mathrm{t} + 3n_\mathrm{h}$, with the type of the ejected nucleon chosen at random with probabilities proportional to the proton and neutron numbers of the parent nucleus, and
        \item alpha particle ejection, with cross section
        $$\sigma_\alpha = \sum_\text{channels} n_\alpha \sigma_{n_\mathrm{n}n_\mathrm{p}n_\mathrm{d}n_\mathrm{t}n_\mathrm{h}n_\alpha}
        = \langle n_\alpha \rangle \sigma_\text{tot}.$$
    \end{itemize}
    where $\sigma_{n_\mathrm{n}n_\mathrm{p}n_\mathrm{d}n_\mathrm{t}n_\mathrm{h}n_\alpha}$ is the
    exclusive cross section for ejecting $n_\mathrm{n}$ neutrons,
    $n_\mathrm{p}$ protons, ..., and $n_\alpha$ $\alpha$-particles as computed by TALYS.
    Finally, for faster computation of the interaction rates, $\sigma_N$ and $\sigma_\alpha$ are fitted by a Gaussian for photon energies between the threshold energy and 30~MeV, and by a constant between 30~and 150~MeV.

\section{Comparisons\label{sec.comparisons}}
In this section outputs of the simulation codes are compared. The comparisons are done for the case of pure composition at injection, with separate discussions for the case of protons (section~\ref{sec.protons}) and nuclei (section~\ref{sec.nuclei}) at the sources.
For each case study, several scenarios are shown through comparisons of observables such as the flux at Earth and, in the case of nuclei injection, the average and variance of the logarithm of the mass. The observables are computed in $\log_{10}(E/\mathrm{eV})$ bins of width~0.1, from 17.5 to 20.5. The uncertainties in the choice of models are studied by computing $(J_i - J_j)/((J_i + J_j)/2)$ for the energy spectrum and $\langle \ln A \rangle_i - \langle \ln A \rangle_j$ and $\sigma^2(\ln A)_i - \sigma^2(\ln A)_j$ for the composition observables. These differences are presented together with the statistical uncertainties in the Auger data for the energy spectrum~\cite{ThePierreAuger:2013eja} and for the composition observables~\cite{Aab:2014kda}, the latter using the EPOS-LHC~\cite{Pierog:2013ria} hadronic interaction model.

\subsection{Propagation of protons}\label{sec.protons}
In this section we compare the observed fluxes of protons, injected with a power law spectrum with spectral index $\gamma = 2.5$ up to a maximum energy of $10^{22.5}$\,eV, in order to assess the similarity between the simulation results for a wide range of initial energies.
The source luminosity per unit of comoving volume is proportional to $(1+z)^3$ in the redshift range $0<z<2.5$.

\subsubsection{Stochasticity of pion production}
In order to study the effect of statistical fluctuations in pion production interactions,
we compare {\it SimProp} results with the option {\tt -S 0}, where pion production is treated deterministically according to the CEL approximation, which has been frequently used in the literature (e.g.~in ref.~\cite{Berezinsky:2002nc}), and with the option {\tt -S 1}, where it is treated as a discrete interaction with a stochastically sampled interaction point and
energy loss. Here, the pion production is computed in the CMB only.

\begin{figure}[t]
    \centering
    \includegraphics[width=0.495\textwidth]{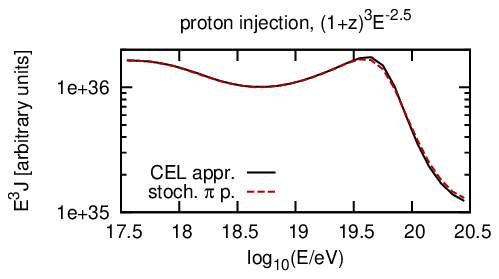}
    \includegraphics[width=0.495\textwidth]{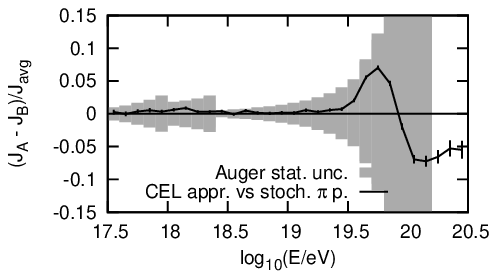}
    \caption{Comparison of observed proton fluxes simulated with pion production treated deterministically and stochastically.}
    \label{fig:stoc}
\end{figure}

The results are shown in figure~\ref{fig:stoc}. The two spectra are very similar,
except that the peak at $10^{19.6}$~eV is broader in the stochastic simulation.
The differences between the two fluxes are less than 10\% at all energies and
are only sizeable at $E \gtrsim 10^{19.5}$~eV, where existing measurements of
the UHECR spectrum have large statistical uncertainties.

\subsubsection{Pion production on the EBL}
In order to study the effect of pion production on the extragalactic background light we compare the proton fluxes computed by {\it SimProp} taking into account the CMB and one of the following EBL models: Stecker et al.~\cite{Stecker:2005qs,Stecker:2006eh}, Kneiske et al.~\cite{Kneiske:2003tx}, Gilmore et al.~\cite{Gilmore:2011ks}, and Dom\'inguez et al.~\cite{Dominguez:2010bv}. For a discussion and comparison of these models see appendix~\ref{sec.EBLmodels}.

\begin{figure}[t]
    \centering
    \includegraphics[width=.495\textwidth]{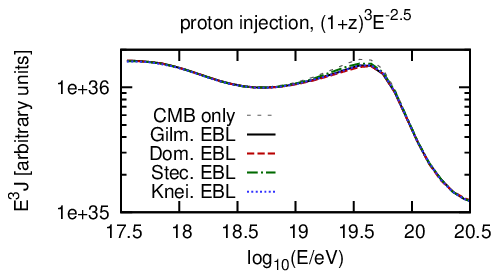}
    \includegraphics[width=.495\textwidth]{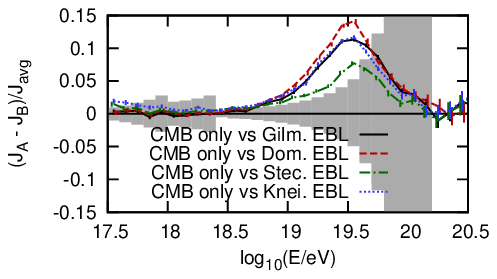}
    \includegraphics[width=.495\textwidth]{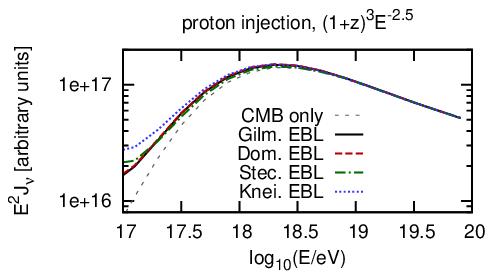}
    \includegraphics[width=.495\textwidth]{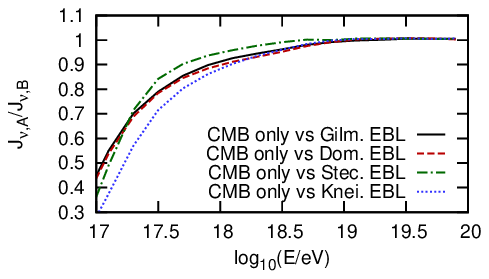}
    \caption{Comparison of observed proton fluxes (top) and cosmogenic neutrino fluxes (bottom) for different EBL models (Stecker, Kneiske, Dom\'inguez and Gilmore), and for neglecting interactions on the EBL.} 
    \label{fig:protonEBL}\label{fig:neutrinoEBL}
\end{figure}

The resulting fluxes are shown in figure~\ref{fig:protonEBL}. Neglecting pion production on the EBL results in more than 10\% higher cosmic ray fluxes at $10^{19.5}$\,eV, but the difference is already visible at energies around $10^{19}$\,eV where the experimental data have small statistical uncertainties.
The difference between EBL models is smaller, but is still at the level of the measured statistical uncertainties at energies around $10^{19.5}$\,eV.

Neutrino fluxes at Earth from the decay of pions and neutrons produced in the same scenario are shown in figure~\ref{fig:neutrinoEBL}, both taking into account the CMB and the various EBL models, as well as the CMB only.
It can be seen that even though the effect of EBL pion production on the proton flux is relatively small, it provides the majority of low energy ($E \lesssim 10^{17.1}$~eV) cosmogenic neutrinos.

\subsubsection{Effect of different simulation codes: {\it SimProp} {\it vs} CRPropa}
We compare {\it SimProp} and CRPropa using the same EBL model (Gilmore et al.~\cite{Dominguez:2010bv}), in order to investigate the effect of different propagation algorithms.
In particular, the two codes use different step lengths for numerical integrations.
In CRPropa the redshift dependence of the interaction rates for interactions with the EBL are approximated through a global scaling factor, whereas in {\it SimProp} rates are calculated at each redshift. In {\it SimProp} all photohadronic processes are approximated as single pion production isotropic in the center of mass frame, and with branching ratios from isospin invariance; neutron decay is treated as instantaneous. CRPropa treats photopion production of protons and neutrons differently, using the SOPHIA code to compute the energy dependent branching ratios and energy losses.

\begin{figure}[t]
    \centering
    \includegraphics[width=0.495\textwidth]{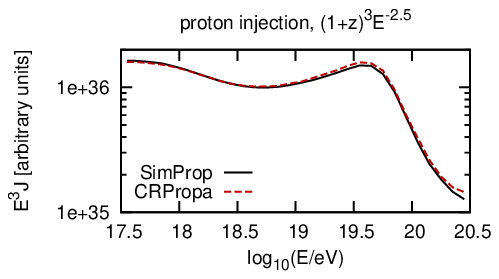}
    \includegraphics[width=0.495\textwidth]{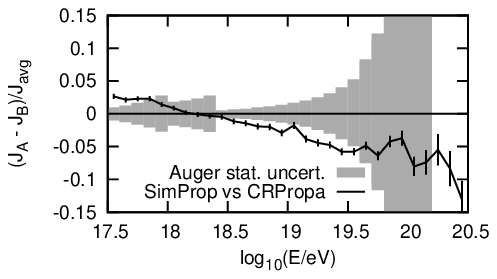}
    \caption{Comparison between proton fluxes at Earth simulated using {\it SimProp}~v2r3 and CRPropa~3.}
    \label{fig:protonsSPvsCRP}
\end{figure}

The simulated proton fluxes for both codes are shown in figure~\ref{fig:protonsSPvsCRP}.
The differences are small ($\lesssim 10\%$), although they systematically depend on the energy, decreasing by about $4\%$ per energy decade. This results in a steeper spectrum for {\it SimProp} than for CRPropa, corresponding to a variation in the spectral index of about~$0.02$.

\subsection{Propagation of nuclei}\label{sec.nuclei}
In this section we study the uncertainties related to the propagation of nuclei.
The general scenario consists of identical sources emitting nuclei with a power law spectrum with rigidity-dependent cutoff $dN/dE \propto E_\text{inj}^{-\gamma}\exp(-E_\text{inj}/ZR_\text{cut})$. The sources are homogeneously distributed in comoving volume, and the redshift range used here is $0<z<1$.
We consider two representative primary nuclides, nitrogen-14 and iron-56, and two representative injection spectra with associated rigidity-dependent cutoffs, $\gamma=2, R_\text{cut}=10^{20}$~V (``soft'') and $\gamma=1, R_\text{cut}=5\times10^{18}$~V (``hard''). (The soft injection is inspired by the best fit of the Auger spectrum and composition results reported in~\cite{Taylor:2011ta} in the case of pure iron injection.  Hard injection of intermediate nuclei is found to be the best fit to Auger data from several authors~\cite{Allard:2011aa,Aloisio:2013hya,Taylor:2013gga} for an arbitrary mixed composition at the source.)
We have simulated the resulting observables for all combinations of injection characteristics (hard iron, soft iron, hard nitrogen, soft nitrogen), for every propagation model studied.
In the following, we show a selection of the investigated source characteristics and briefly describe qualitative differences for the cases that are not shown.

\subsubsection{Effect of pion production by nuclei}
In order to study the importance of pion production by nuclei, we compare fluxes and compositions computed with {\it SimProp} for the provided options of neglecting and of considering pion production on bound nucleons. In the former (option {\tt -S -1}), pion production for protons is approximated as a CEL process, whereas in the latter (option {\tt -S 1}) it is treated stochastically. In both cases, the Gilmore et al.~\cite{Gilmore:2011ks} EBL model and the PSB photodisintegration model are used.

\begin{figure}[t]
    \centering
    \includegraphics[width=0.325\textwidth]{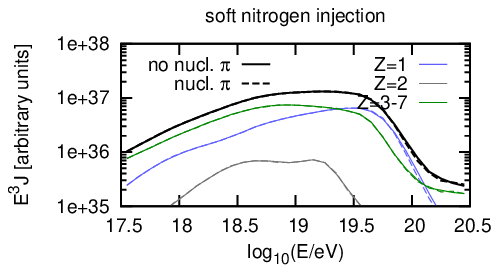}
    \includegraphics[width=0.325\textwidth]{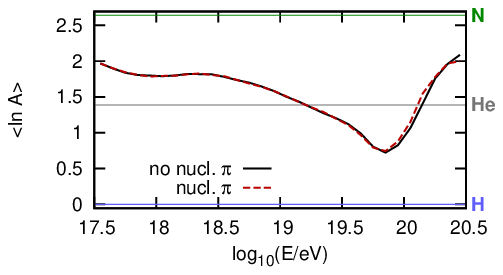}
    \includegraphics[width=0.325\textwidth]{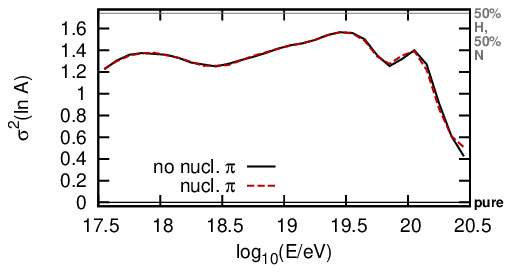}
    \includegraphics[width=0.325\textwidth]{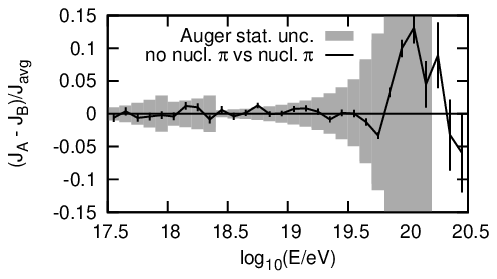}
    \includegraphics[width=0.325\textwidth]{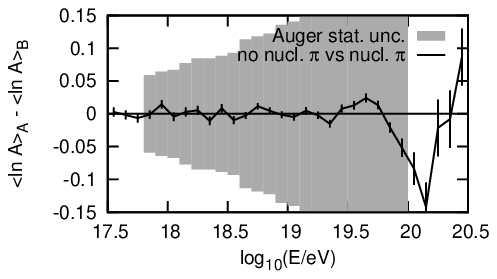}
    \includegraphics[width=0.325\textwidth]{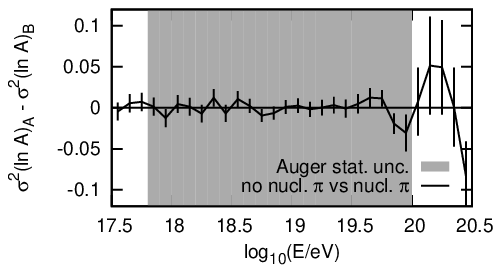}
    \caption{Effect of pion production on nuclei for soft nitrogen injection.} 
    \label{fig:pion_softN}
\end{figure}

The results are shown in figure~\ref{fig:pion_softN} for the case of soft nitrogen injection.
It can be seen that the effect of pion production by nitrogen nuclei is negligible, except at energies $E > 10^{19.7}\,$eV where the available measurements have large statistical uncertainties. This effect is even smaller in the case of iron primaries, since the energy threshold for pion production increases with nuclear mass, and in the case of hard injection, as the lower injection cutoff implies fewer primaries above the threshold.

\subsubsection{Effect of different EBL models}
\begin{figure}[t]
    \centering
    \includegraphics[width=0.325\textwidth]{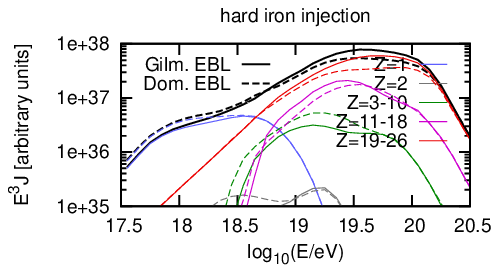}
    \includegraphics[width=0.325\textwidth]{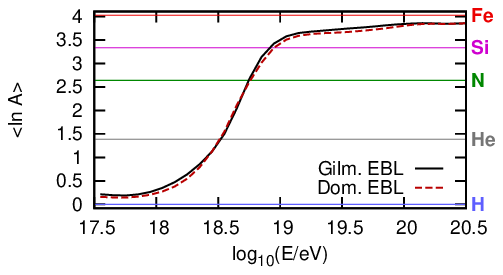}
    \includegraphics[width=0.325\textwidth]{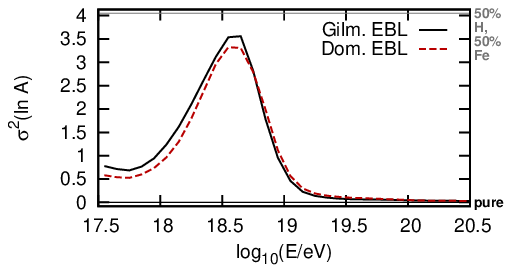}
    \includegraphics[width=0.325\textwidth]{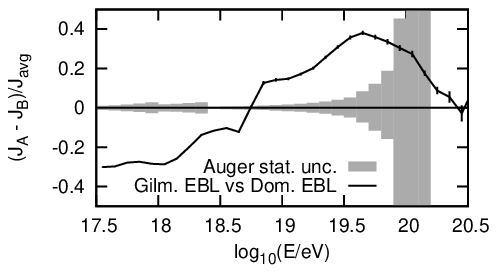}
    \includegraphics[width=0.325\textwidth]{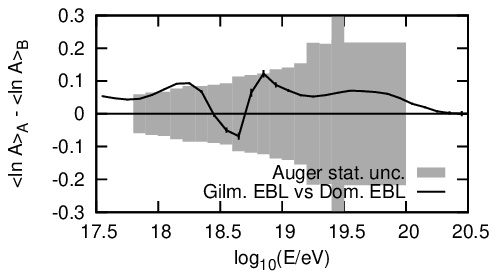}
    \includegraphics[width=0.325\textwidth]{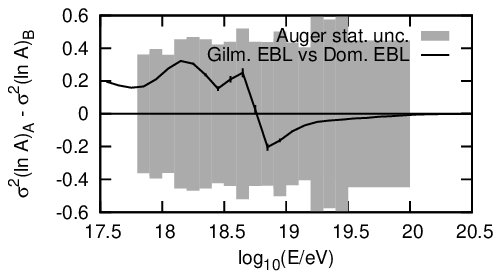}
    \caption{Comparison of EBL models (Gilmore {\it vs} Dom\'inguez) for hard iron injection.} 
    \label{fig:EBL_hardFe}
\end{figure}

\begin{figure}[t]
    \centering
    \includegraphics[width=0.325\textwidth]{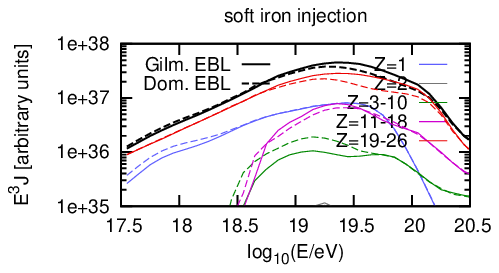}
    \includegraphics[width=0.325\textwidth]{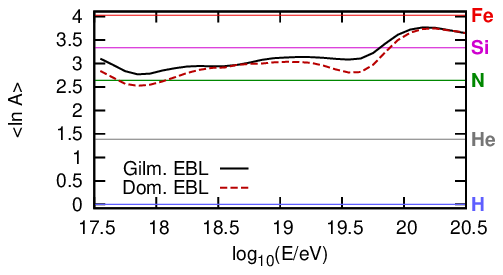}
    \includegraphics[width=0.325\textwidth]{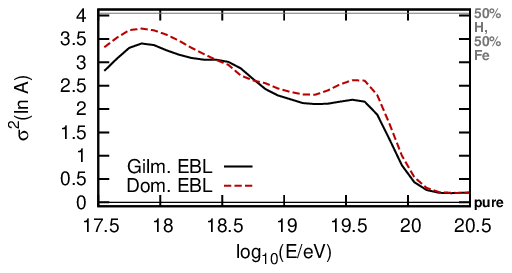}
    \includegraphics[width=0.325\textwidth]{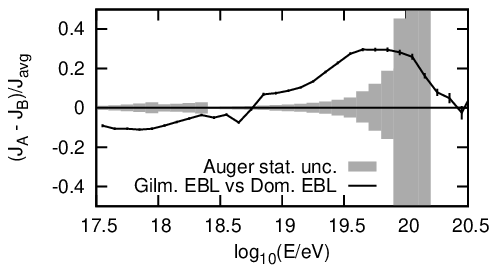}
    \includegraphics[width=0.325\textwidth]{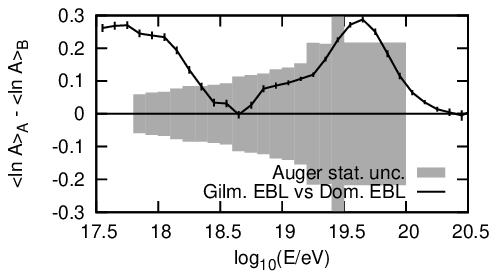}
    \includegraphics[width=0.325\textwidth]{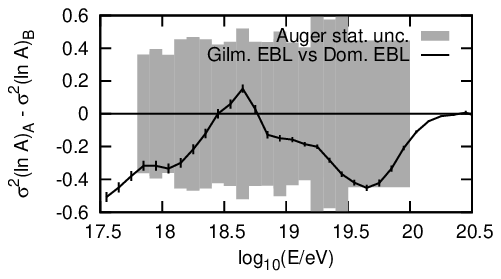}
    \caption{Comparison of EBL models (Gilmore {\it vs} Dom\'inguez) for soft iron injection.} 
    \label{fig:EBL_softFe}
\end{figure}

To study the impact of different EBL models on the photodisintegration of nuclei, we have used {\it SimProp} with the two most up-to-date EBL models, namely Gilmore {\it et al.} and Dom\'inguez {\it et al.}. The differences between these models are discussed in appendix~\ref{sec.EBLmodels}. In both cases the PSB photodisintegration model is used. The results are shown in figure~\ref{fig:EBL_hardFe} for hard iron injection and in figure~\ref{fig:EBL_softFe} for soft iron injection.

The differences in the total energy spectra are significant, often exceeding 10\%, resulting in softer spectra at Earth for the stronger Dom\'inguez EBL model than for the Gilmore model.
The differences in the spectrum are more clearly visible in the hard injection scenarios (more than 40\% at $E \sim 10^{19.3}$\,eV), mainly due to different predicted numbers of low energy secondaries, while in soft injection scenarios low energy secondary protons are subdominant with respect to primary nuclei. Conversely, the differences in the average logarithmic mass are larger for hard injection, because the composition is dominated by secondary protons at low energy and primary nuclei at high energy with either model, whereas with soft injection the composition is more mixed and more model-dependent.

The differences in partial spectra are mostly visible for low-energy intermediate mass secondaries of iron, because they are produced via repeated disintegration by the EBL.
At high energies the partial spectra are in good agreement, because high-energy nuclei are mainly disintegrated by the CMB. The cases of nitrogen injection give very similar results for both hard and soft injection.

\subsubsection{Effect of photodisintegration cross sections I: PSB {\it vs} TALYS}
To investigate the effect of different photodisintegration cross section models, we compare {\it SimProp} simulations with both the PSB and the TALYS photodisintegration models. In both cases we used the Gilmore {\it et al.} EBL model.

\begin{figure}[t]
    \centering
    \includegraphics[width=0.325\textwidth]{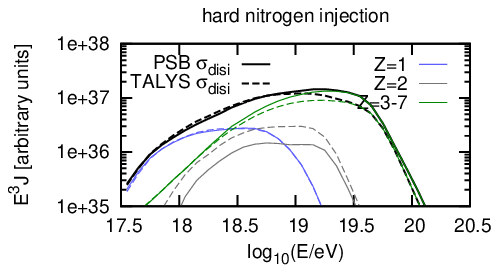}
    \includegraphics[width=0.325\textwidth]{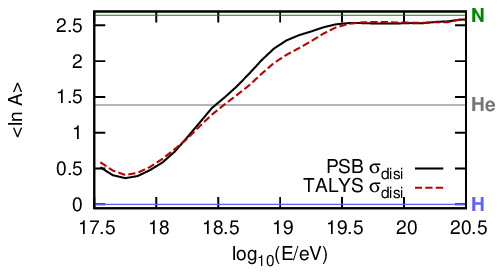}
    \includegraphics[width=0.325\textwidth]{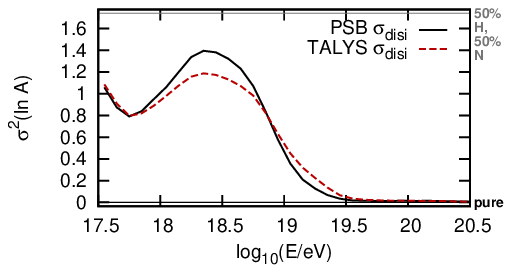}
    \includegraphics[width=0.325\textwidth]{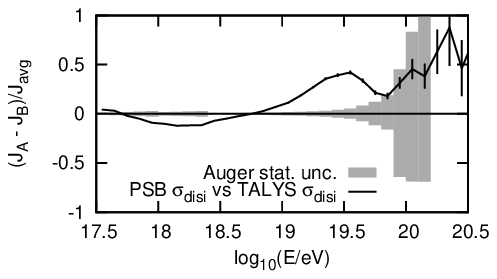}
    \includegraphics[width=0.325\textwidth]{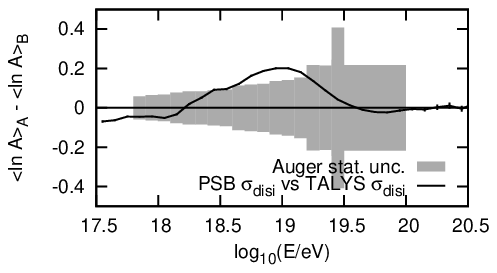}
    \includegraphics[width=0.325\textwidth]{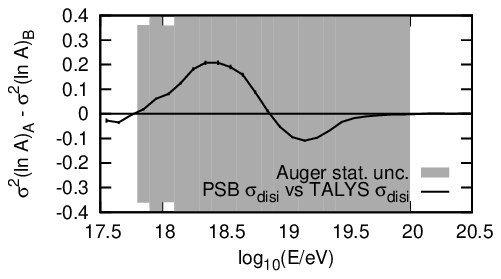}
    \caption{Comparison of PSB and TALYS photodisintegration models for hard nitrogen injection.} 
    \label{fig:sigmaI_hardN}
\end{figure}

The results are shown in figure~\ref{fig:sigmaI_hardN} for the case of hard nitrogen injection.
The main difference between the PSB and TALYS models of photodisintegration cross sections is that only the latter includes channels where $\alpha$-particles, rather than single nucleons, are ejected. The effect of these channels can be seen in the spectra at Earth, where the helium flux using TALYS largely exceeds the one obtained using PSB. In the case of soft nitrogen injection, differences in the total energy spectrum are about half of those for hard injection (because in that case secondaries are subdominant with respect to primaries even at low energies) and those in the average logarithmic mass are similar. The cases of iron injections result in very small differences (because channels ejecting $\alpha$-particles are even more disfavoured with respect to those ejecting single nucleons than in the case of nitrogen).

\subsubsection{Effect of photodisintegration cross sections II: TALYS {\it vs} Kossov}
A third model for the photodisintegration cross sections is provided by the parametrization of Kossov~\cite{kossov2002}, which is used in the GEANT4 code. Since the model does not parametrize partial cross sections, its total cross sections can be used in CRPropa in combination with the branching ratios from TALYS.
We compare the results of CRPropa simulations using the TALYS and Kossov cross sections. In both cases the Dominguez et al. EBL model is used.

\begin{figure}[t]
    \centering
    \includegraphics[width=0.325\textwidth]{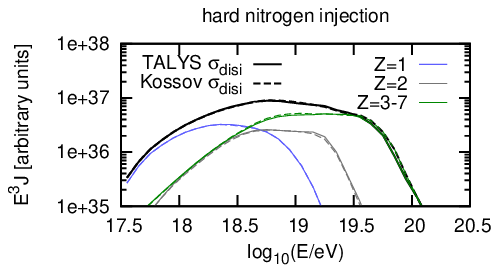}
    \includegraphics[width=0.325\textwidth]{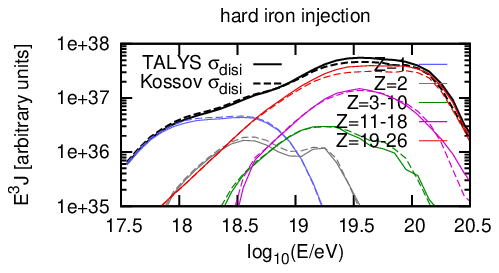}\\
    \includegraphics[width=0.325\textwidth]{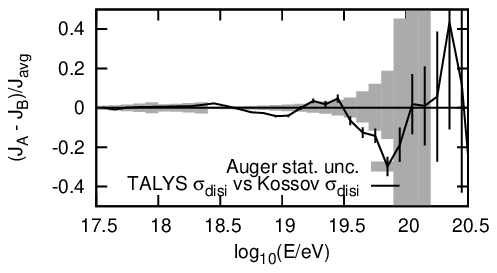}
    \includegraphics[width=0.325\textwidth]{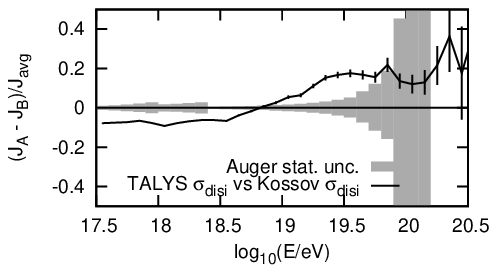}
    \caption{Comparison of TALYS and Kossov photodisintegration cross sections for hard injection of nitrogen (left) and iron (right) nuclei.} 
    \label{fig:sigmaII_1st}
\end{figure}

The resulting fluxes are shown in figure~\ref{fig:sigmaII_1st} for hard injection of nitrogen and iron nuclei.
Using the total cross sections from Kossov results in a higher level of photodisintegration for iron nuclei compared to TALYS, resulting in a difference in the spectrum of around 20\% for $E \gtrsim 10^{19}$\,eV.
For nitrogen the difference is much smaller in the energy range with low statistical uncertainty.
The results are similar for both hard and soft injection scenarios.
In our simulations the differences between TALYS and Kossov are small, compared to that between TALYS and PSB, or between using different EBL models.
This is due to similar total cross sections and because the same branching ratios (from TALYS) are used in both cases.

\subsubsection{Effect of photodisintegration cross sections III: TALYS with rescaled $\sigma_\alpha$}
In this section we investigate the influence of photodisintegration channels where $\alpha$-particles are ejected on the spectrum and composition observables. To this end, we ran {\it SimProp} simulations using TALYS photodisintegration where all values of $\sigma_\alpha$ (defined in section~\ref{sec:v2r3}) were scaled by a factor $a_{\alpha}=1.0$ (unscaled), 0.3, 0.1, and 0.0 ($\alpha$ ejection disabled). In each case the Gilmore EBL model was used.

\begin{figure}[t]
    \centering
    \includegraphics[width=0.325\textwidth]{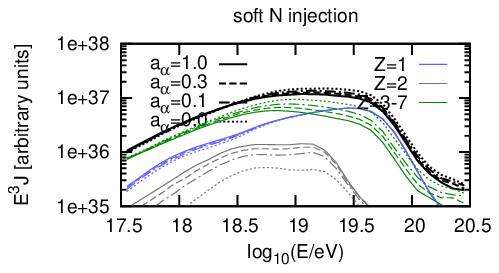}
    \includegraphics[width=0.325\textwidth]{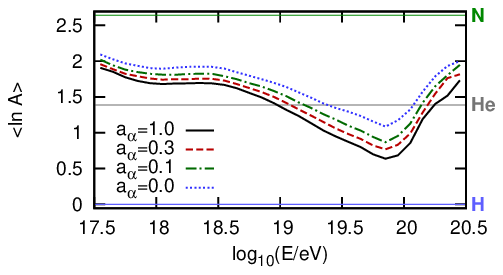}
    \includegraphics[width=0.325\textwidth]{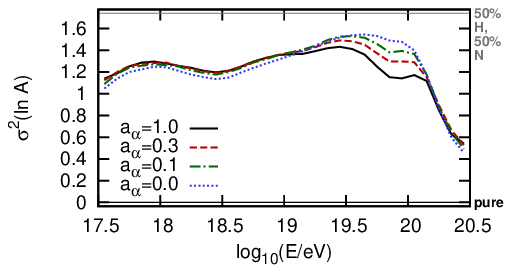}
    \includegraphics[width=0.325\textwidth]{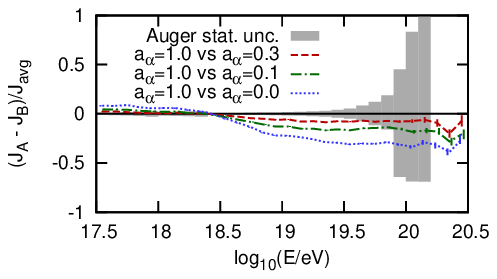}
    \includegraphics[width=0.325\textwidth]{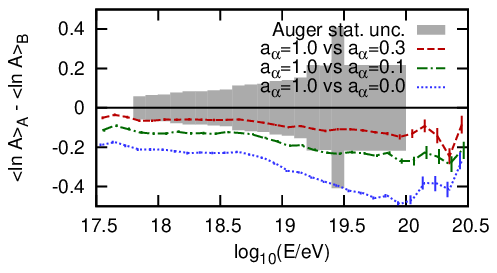}
    \includegraphics[width=0.325\textwidth]{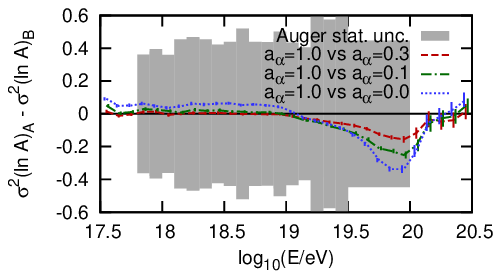}
    \caption{Effect of scaling of TALYS cross sections for $\alpha$-particle ejection for soft nitrogen injection.} 
    \label{fig:alpha_softN}
\end{figure}

The reason for this kind of analysis is the lack of cross section measurements for the $\alpha$-particle ejection: in the data sets used in the TALYS code the only measurements for this photodisintegration channel (for nuclei with $10 \leq A \leq 56$ and photons below 30 MeV in the nucleus rest frame) are the ones of $\rm {}^{12}C$ ($\rm {}^{12}C + \gamma \rightarrow 3 \alpha$, $\rm {}^{12}C + \gamma \rightarrow p + \alpha + {}^{7}Li$) and $\rm {}^{16}O$ ($\rm {}^{16}O + \gamma \rightarrow 4 \alpha$). Moreover, TALYS seems to overpredict these measurements, as shown in appendix~\ref{sec.GDRparams}.
It is thus worthwhile to understand what the impact of these uncertainties is on the observables.
Results are shown in figure~\ref{fig:alpha_softN} and following.

The effects of changing this poorly known quantity on the energy spectrum in the case of hard nitrogen injection can be very large, over 50\%, resulting in softer spectra at Earth the larger $a_\alpha$ is. In the case of soft nitrogen injection these differences are about half as large, as there are more primaries at low energy and more secondaries at high energy than for hard injection. Conversely, the effects of this scaling on the average logarithmic mass in the case of soft nitrogen injection are large at all energies, whereas those for hard nitrogen injection are similar at low energies but negligible at high energies, where the composition is almost purely primary nitrogen regardless of the value of $a_\alpha$. In the case of iron injection, the effects of the scaling are smaller, because for heavy nuclei $\alpha$-particle ejection is strongly disfavored.

\begin{figure}[t]
    \centering
    \includegraphics[width=0.325\textwidth]{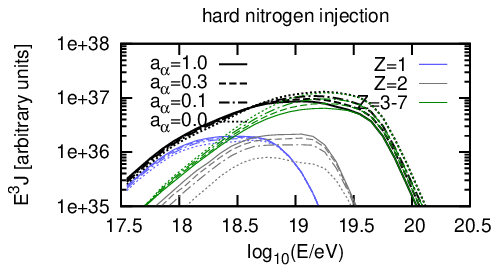}
    \includegraphics[width=0.325\textwidth]{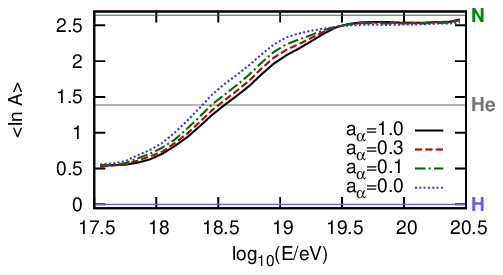}
    \includegraphics[width=0.325\textwidth]{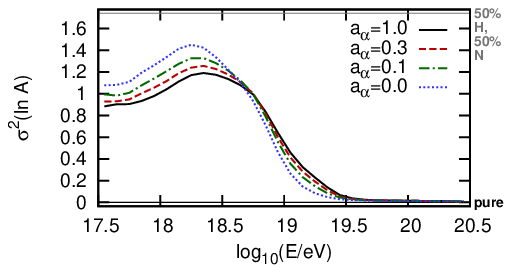}
    \includegraphics[width=0.325\textwidth]{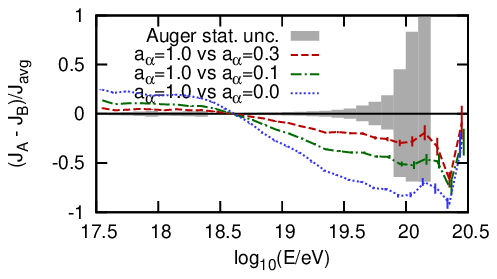}
    \includegraphics[width=0.325\textwidth]{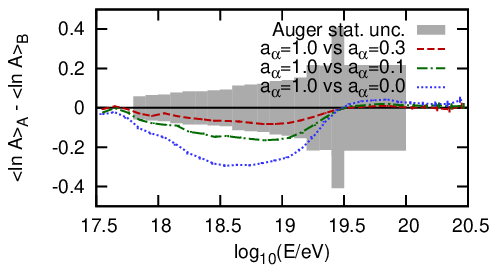}
    \includegraphics[width=0.325\textwidth]{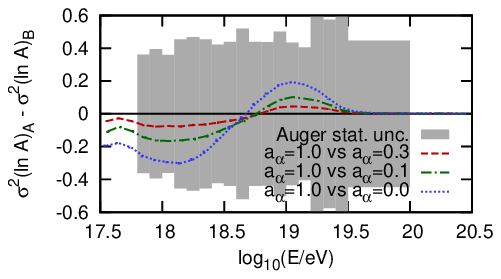}
    \caption{Effect of scaling of TALYS cross sections for $\alpha$-particle ejection for hard nitrogen injection.} 
    \label{fig:alpha_hardN}
\end{figure}

\begin{figure}[t]
    \centering
    \includegraphics[width=0.325\textwidth]{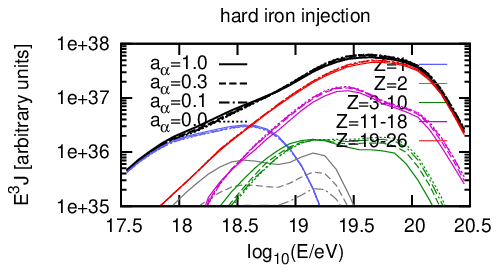}
    \includegraphics[width=0.325\textwidth]{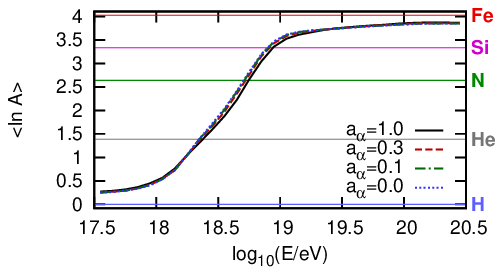}
    \includegraphics[width=0.325\textwidth]{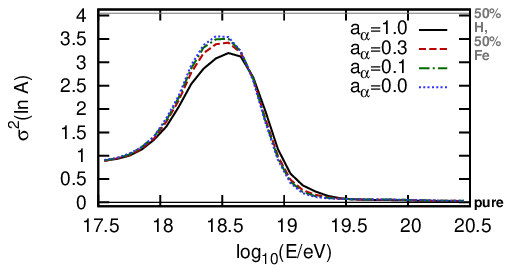}
    \includegraphics[width=0.325\textwidth]{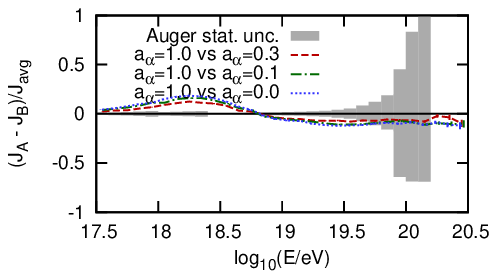}
    \includegraphics[width=0.325\textwidth]{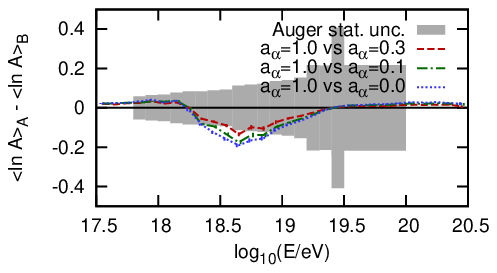}
    \includegraphics[width=0.325\textwidth]{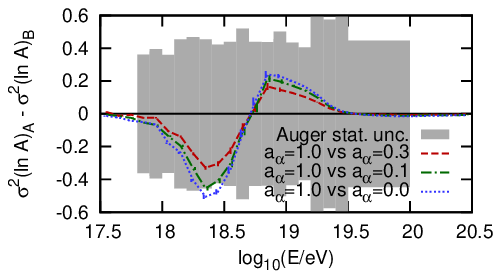}
    \caption{Effect of scaling of TALYS cross sections for $\alpha$-particle ejection for hard iron injection.} 
    \label{fig:alpha_hardFe}
\end{figure}

\subsubsection{Effect of different propagation codes: {\it SimProp} {\it vs} CRPropa}
In order to study the effect of the different simulation algorithms on the propagation of nuclei, we compare fluxes and composition observables computed by {\it SimProp} and CRPropa, both using the Gilmore EBL model and TALYS photodisintegration cross sections.

\begin{figure}[t]
    \centering
    \includegraphics[width=0.325\textwidth]{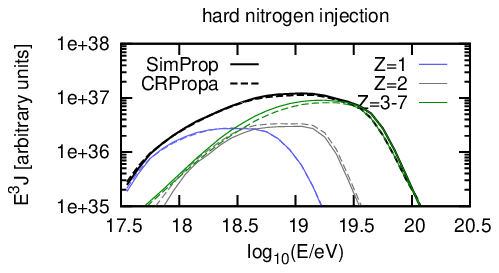}
    \includegraphics[width=0.325\textwidth]{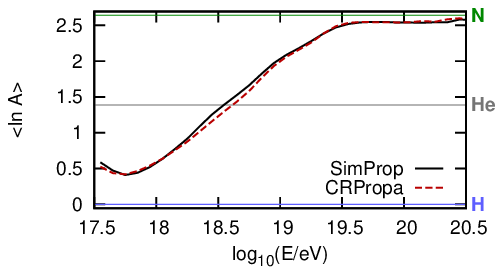}
    \includegraphics[width=0.325\textwidth]{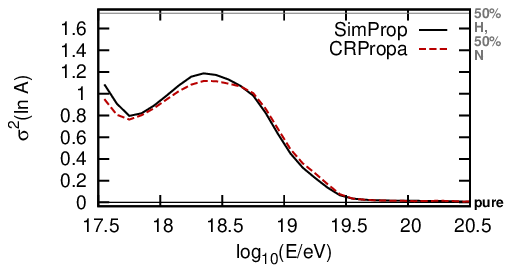}
    \includegraphics[width=0.325\textwidth]{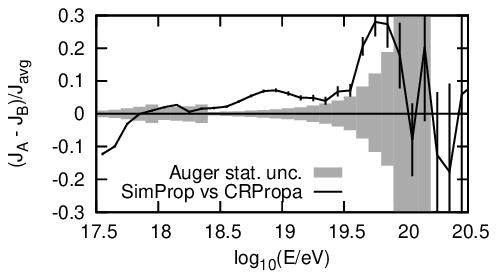}
    \includegraphics[width=0.325\textwidth]{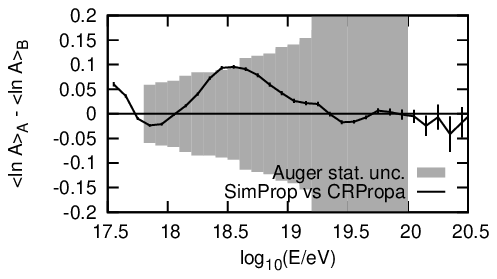}
    \includegraphics[width=0.325\textwidth]{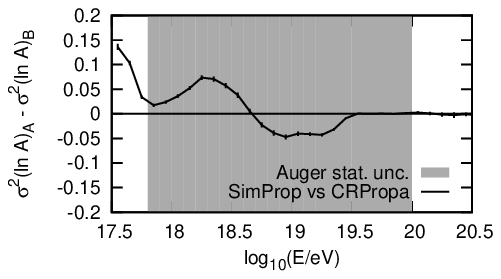}
    \caption{Effect of different propagation codes for hard nitrogen injection} 
    \label{fig:code_hardN}
\end{figure}

The results for hard nitrogen injection are shown in figure~\ref{fig:code_hardN}. It can be seen that once the two codes are used with the same models for the EBL spectrum and photodisintegration cross sections, the remaining differences due to the different approximations used in the algorithms are small (of the order of 10\% or less except at the highest energies), although larger than the statistical uncertainties on the energy spectrum. The cases of soft nitrogen injection and iron injections result in similar or smaller differences.

Possible reasons for the remaining differences include the simplified treatment of photodisintegration in {\it SimProp} described in section~\ref{sec:v2r3}, the simplified redshift-dependence of photodisintegration on the EBL in CRPropa, described in section~\ref{sec:CRP}, or the different cross sections for light nuclei (cross sections as listed in section~\ref{sec:CRP} for $A<12$ in CRPropa, PSB cross sections for $A<5$ in {\it SimProp}).

\section{Discussion\label{sec.discussion}}


We studied the propagation of protons and its effect on energy spectra at Earth with fixed assumptions for the injection spectrum at the source. An analysis concerning the propagation of nuclei has been done for two representative primary nuclides and two representative injection spectra. One of the main purposes of this study was to quantitatively estimate the impact of different propagation models in the observables. The observed differences in the spectrum and composition at Earth could imply variations in the source models predicted by several authors such as the ones proposed in refs.~\cite{Allard:2011aa,Aloisio:2013hya,Taylor:2013gga}. The differences in the calculated observables are shown together with the Auger statistical uncertainties in order to assess whether they can possibly have sizeable effects on the results of fits to Auger data.

The effect of different implementations of photopion production has been tested. We have quantified how the approximation of photopion production as a continuous energy loss process, as done e.g. in refs.~\cite{Berezinsky:2002nc,Aloisio:2006wv}, affects the measured spectrum and mass composition, compared to the stochastic treatment. There are visible differences at energies $\gtrsim 10^{19.5}\,$eV, though they are below 10\%. These differences tend to decrease for heavier nuclei, due to the dependence of the energy threshold for photopion production on the mass of the particle.


Differences in the spectral density and redshift evolution of the EBL models are expected to have an impact on the studied observables. Most of the latest EBL models agree on the overall shape of the spectral density, particularly at $z=0$, and in the ultraviolet and visible regions. Nevertheless, there are some differences that impact the propagation of UHECRs, which is mainly affected by the infrared region, affecting both the spectrum and mass composition. These small differences tend to increase with redshift, thus impacting the study of cosmogenic neutrino fluxes. As a consequence, one needs to constrain the uncertainties due to the choice of EBL models by using, for example, multiple models or the available upper and lower bounds (see ref.~\cite{Dwek:2012nb} for a review).

In the case of proton injection, among the EBL models considered, the largest bump in the UHE proton spectrum at Earth, apart from the one due to the CMB, is obtained using the Stecker et al. EBL model~\cite{Stecker:2006eh} whose intensity in the infrared peak is lower compared to the other models studied (see appendix~\ref{sec.EBLmodels}, fig.~\ref{fig:ebl}). The choice of the EBL model does not significantly affect the spectrum measured at Earth in this particular case, since the CMB dominates over the EBL. However, it is extremely important for studies involving cosmogenic neutrinos at PeV energies, since it is directly involved in their production. Nevertheless, one should bear in mind that although the flux of neutrinos produced due to proton interactions with photon fields can be affected by this choice, this uncertainty is smaller than the uncertainties in the cosmological evolution of sources \cite{Aloisio:2015ega}. 
In the case of nuclei, due to the giant dipole resonance, at energies $\sim 10\,$EeV photodisintegration can occur via interaction with photons with energies $\sim 10-100\,$meV in the laboratory frame. For this reason, the EBL plays a fundamental role in the propagation of nuclei.
In the particular case of the Gilmore et al.~\cite{Gilmore:2011ks} and Dom\'inguez et al.~\cite{Dominguez:2010bv} EBL models the differences in the energy distribution of photons can result in observed differences of up to 40\% in the spectrum, and 30\% in the average logarithm of the mass number, exceeding statistical uncertainties from Auger, thus being non-negligible. As expected, differences are visible up to $\sim 10^{20.2}\,$eV in the case of iron injection, while in the case of nitrogen injection they are visible up to $\sim 10^{19.7}\,$eV, due to the lower value of the threshold for the photodisintegration in the CMB.

In ref.~\cite{Hooper:2006tn} the authors compare three EBL models, namely the ones from refs.~\cite{Malkan:2000gu,Aharonian:2003tr,Franceschini:2001yg}, and claim that there are considerable differences among them only for particularly heavy nuclei. Their conclusion is based on the comparison of energy loss lengths for photodisintegrations assuming these EBL models at a fixed redshift ($z=0$). Moreover, the EBL models compared in ref.~\cite{Hooper:2006tn} appear to have a larger impact at the lowest energies, where indeed the energy loss length of photodisintegration in the EBL, which at these energies is mainly due to the UV peak of the EBL density distribution, is much larger than the one for adiabatic losses. The IR peak of its distribution is then the main contributor to photodisintegration in the EBL. As shown in fig.~\ref{fig:ebl}, the discrepancies among different models are larger for larger wavelengths, making a discussion of their influence on the observables necessary. Furthermore, the spectral energy distribution of the EBL evolves with redshift, making the cosmic ray spectrum a better observable to study the effects of EBL models than the energy loss length only.


Different approaches in the propagation algorithms, as implemented in CRPropa and {\it SimProp}, have been investigated. In the case of protons the differences do exceed the statistical uncertainties of Auger in the energy range between $10^{18.5}\,$eV and $10^{19.5}\,$eV, although they are always smaller than 10\%. 
The different approximations adopted for the treatment of photopion production are likely to be the main reason for the discrepancies.
We have also compared the two simulation codes for the case of nuclei, adopting the same EBL model and photodisintegration cross sections. The differences are of the order of 10\%, except for energies above $10^{20}\,$eV, in which case this number can reach 15\%. Possible reasons for these differences are the simplified treatment of the photodisintegration in {\it SimProp}, the simplified treatment of the EBL redshift evolution in CRPropa, or the different cross sections used for light nuclei, besides the differences in the treatment of protons, which may be produced via photonuclear interactions in the case of primary nuclei.

A substantial part of the present work is dedicated to the discussion of the impact of photodisintegration cross sections on observables. The default version of {\it SimProp} uses the PSB photodisintegration model, which does not include channels whereby $\alpha$ particles are ejected, as TALYS does. The consequences of this simplification can be easily seen in the measured flux of individual species: there is a significant enhancement in the flux of helium nuclei and a reduction in the flux of intermediate mass nuclei such as nitrogen. This results in a decrease of $\langle \ln A \rangle$ in the energy range between $\sim 10^{18.5}\,$eV and $10^{19.5}\,$eV. The lack of measurements of photonuclear cross sections in the data sets used by TALYS, and the fact that TALYS overpredicts some of the measured cross sections, motivated the study of the effect of rescaling the $\alpha$ production rates. Moreover, the uncertainties associated with $\langle \ln A \rangle$ by Auger are, in principle, small enough to allow us to distinguish between protons and helium nuclei. Therefore, it is important that UHECR propagation simulations accurately predict the number of protons and $\alpha$-particles reaching Earth. The ejection of $\alpha$ particles is less frequent in the case of iron than in the case of nitrogen, and therefore the effects of this scaling in the iron injection scenarios are smaller than in the nitrogen cases. 

The parametrization of cross sections used in GEANT 4~\cite{kossov2002} together with the branching ratios from TALYS is compared with the TALYS cross sections in CRPropa. Differences are small compared to those arising from using different EBL models, and from including the $\alpha$-particle ejection.

As a general remark, the differences in the spectra are often more visible in the hard injection scenario ($\gamma=1, R_\text{cut}=5\times 10^{18}$ V) than in the soft one ($\gamma=2, R_\text{cut}=10^{20}$ V), because they are mainly driven by the ratio of primary to (low energy) secondaries, which is smaller in the soft than in the hard injection scenarios. On the other hand, considering the composition observables only ($\langle \ln A \rangle$ and $\sigma(\ln A)$), the differences are smaller for harder injection spectra. This can be interpreted in terms of the low rigidity cutoff in the hard injection scenario, which implies a virtually negligible contribution of lighter nuclei at the highest energies, allowing only the residual primaries of heavier nuclei to arrive at Earth.

In refs.~\cite{Allard:2011aa,Aloisio:2013hya,Taylor:2013gga} it has been argued that hard spectral indices ($\gamma \lesssim 1.6$) and lower maximal energies ($E_{max} \sim Z \times 5 \times 10^{18}\,$eV) are needed in order to simultaneously fit the spectrum and composition measured by Auger. Assuming a rigidity dependent cutoff for the injection spectrum, the cutoff at such low energies arises from the need to have a vanishing light component at the highest energies. As a consequence, the hard spectral indices are required to obtain the right ratio between residual primaries and lighter masses at Earth. In fact, as we have shown in the plots of the spectra, for sources injecting nuclei with a fixed maximum energy $E_{max}$, the harder the injected spectrum is, the larger are the fluxes of secondary nuclei with respect to the residual primaries. Therefore, a detailed understanding of photonuclear processes is important in order to address this issue, since the flux of different particle species arriving at Earth may be significantly under- or overestimated when adopting different approaches.

Understanding the impact of EBL models is also important. In fact, we have demonstrated that the enhancement of the efficiency of photodisintegration, as it occurs when the $\alpha$-channel is included, has qualitatively the same effect as considering the most intense EBL model. One can then argue that a scenario in which the most intense EBL model is chosen together with a photodisintegration model that includes the $\alpha$-channel will result in an extremely efficient production of secondary nuclei. Taylor et al.~\cite{Taylor:2015rla} assume that the uncertainties in the EBL distribution and in cross sections relevant for nuclei propagation are unlikely to qualitatively impact the conclusions on the fit of the Auger data. Nonetheless, we have quantitatively estimated the influence of some combinations of EBL and photodisintegration models on the observables at Earth, showing that even with the simplified choice of pure injection the differences often exceed the statistical uncertainties of Auger.


\section{Conclusions\label{sec.conclusions}}

In the present work we have discussed how sensitive the UHECR propagation is to uncertainties in the EBL spectrum and in the photodisintegration models. This has been investigated using two different simulation codes, CRPropa and {\it SimProp}, which have been directly compared in order to understand the effect of different computational treatments in the observables at Earth. Our results suggest that uncertainties in the scaling of $\alpha$-channels related to the ejection of $\alpha$-particles is the dominant source of uncertainties amongst all studied parameters, including different EBL models and photodisintegration cross sections.  Also, the energy spectrum at Earth is more sensitive to the uncertainties in propagation in scenarios with hard injection spectra, whereas the measured mass composition is more model-dependent for soft injection spectra.

To summarize, we have shown that different choices of parameters such as photonuclear cross sections, EBL model and computational treatment, can have a considerable impact in UHECR observables such as the spectrum as composition. The present work could be used as a quantitative estimation of uncertainties due to propagation in such interpretations.


\acknowledgments
We thank the Pierre Auger Collaboration for permission to use their data prior to journal publication. The authors would also like to thank TALYS developers A.~Koning and S.~Goriely for their help in understanding the differences between the various versions of TALYS.
The research activity of DB is supported by SdC Progetto Speciale Multiasse 
``La Societ\`a della Conoscenza in Abruzzo'', PO FSE Abruzzo 2007 - 2013.
The work of AvV is supported by the Deutsche Forschungsgemeinschaft through the collaborative research centre SFB 676, by BMBF under grant 05A11GU1, by the "Helmholtz Alliance for Astroparticle Physics (HAP)" funded by the Initiative and Networking Fund of the Helmholtz Association and by the State of Hamburg, through the Collaborative Research program ``Connecting Particles with the Cosmos'' as well as the NWO astroparticle physics grant WARP. RAB acknowledges the financial support from the Forschungs- und Wissenschaftsstiftung Hamburg.
\appendix
\section{Photodisintegration cross sections\label{sec.photodis}}
\label{sec.GDRparams}
One of the most important processes in the propagation of ultra-high energy nuclei through diffuse background radiation is photodisintegration, where a nucleus absorbs a photon and then ejects one or more fragments (most commonly single nucleons, but also $\alpha$-particles or multiple nucleons).
The experimental data about the cross sections of such processes are limited: for many nuclei only measurements of the total photoabsorption cross section and/or of single neutron ejection are available, mainly due to the difficulty in detecting outgoing charged particles. For this reason phenomenological models are needed in order to implement these processes in UHECR propagation simulation codes.

The model proposed by Puget, Stecker and Bredekamp~\cite{Puget:1976nz} includes a restricted list of nuclides (with one isobar for each $A$ from 2 to 4 and from 9 to 56), and approximates the cross sections for one- and two-nucleon ejection for photon energies in the nucleus rest frame $2~{\rm MeV} \le \epsilon' \le 30~{\rm MeV}$ as Gaussians, and cross sections for multi-nucleon ejection for $30~{\rm MeV} \le \epsilon' \le 150~{\rm MeV}$ as constants, with tabulated branching ratios for the possible number of ejected nucleons. The exception is beryllium-9, for which the only photodisintegration channel is into two nuclei of helium-4 and one proton. 
A refinement of this model by Stecker and Salamon~\cite{Stecker:1998ib} uses the kinematic threshold for each process instead of $2~{\rm MeV}$ as the lower limit of integration.
Throughout this work, by PSB model we refer to this refinement.
The PSB model makes no distinction between ejected protons and neutrons; when it is used in {\it SimProp}, the corresponding branching ratios are taken to be proportional to the number of protons and neutrons in the parent nucleus.
Also, channels involving the ejection of fragments other than single nucleons (e.g. $\alpha$-particles) are neglected.

\begin{figure}[t]
    \centering\includegraphics[width=0.5\textwidth]{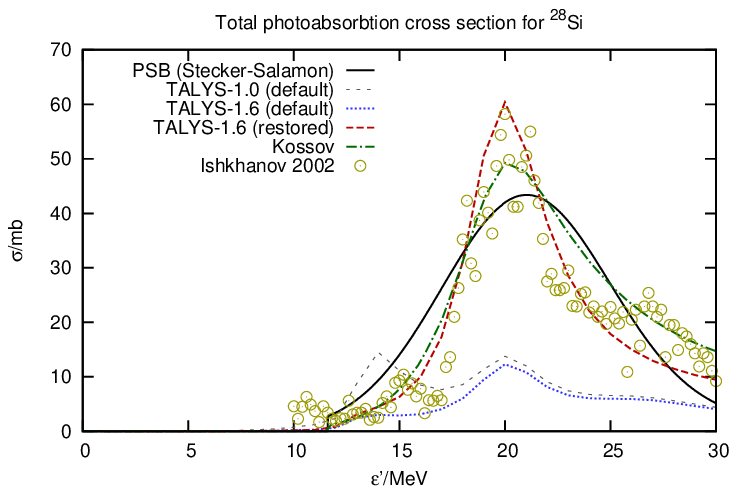}\includegraphics[width=0.5\textwidth]{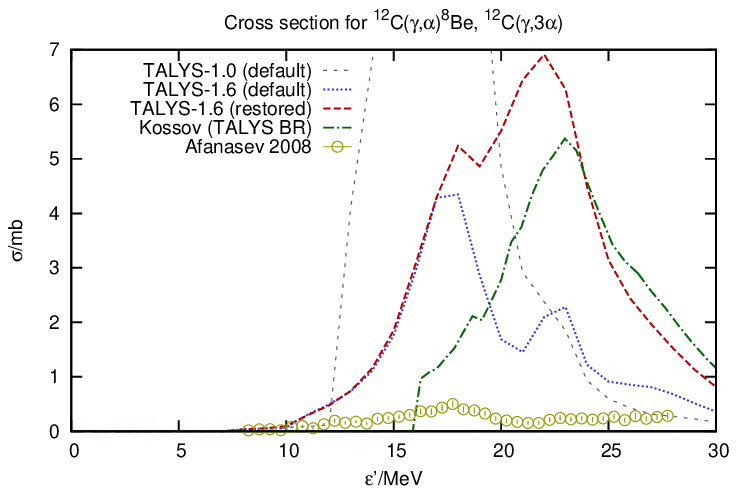}
    \caption{Photodisintegration cross sections for total absorption by silicon-28 (left) and $\alpha$-particle ejection from carbon-12 (right) as predicted by various models. The measured data (yellow circles) are from ref.~\cite{sigmasilicon} and~\cite{sigmaalpha} respectively. TALYS (default) refers to using default TALYS settings, TALYS (restored) to using the GDR parameters listed in table~\ref{tab:GDRparameters}.}
    \label{fig:sigma}
\end{figure}

A more sophisticated model is provided by the nuclear reaction program TALYS~\cite{talys}.
It allows to compute cross sections for all exclusive photodisintegration channels, describing the ejection of protons, neutrons, deuterons, tritons, helium-3 and helium-4 nuclei, and any combinations thereof.
A preliminary version of TALYS was used by Khan et al.~\cite{khan2005} for an exhaustive comparison to the available experimental data.
In their comparison TALYS was used with the giant dipole resonance parameters compiled in the atlas of GDR parameters in ref.~\cite{iaea-tecdoc-1178}.
In contrast, the publicly available versions of TALYS take these parameters from the RIPL-2 database~\cite{iaea-tecdoc-1506} by default.
In this work we make use of the former parameters, if available~\cite{talysmails} (see table~\ref{tab:GDRparameters} for a complete list), as the resulting cross sections (listed as  ``TALYS-1.6 (restored)'' in figure~\ref{fig:sigma}) are in much better agreement with the available measurements.

\begin{table}[t]
\centering
\begin{tabular}{ l || c | c | c | c | c | c | c}
    \hline
    Isotope & $E_0$\,[MeV] & $\sigma_0$\,[mb] & $\Gamma_0$\,[MeV] & $E_1$\,[MeV] & $\sigma_1$\,[mb] & $\Gamma_1$\,[MeV] & Source \\
    \hline
    C-12   &  22.70  &  21.36  &  6.00  &  &  &  &  Atlas \\
    N-14   &  22.50  &  27.00  &  7.00  &  &  &  &  Atlas \\
    O-16   &  22.35  &  30.91  &  6.00  &  &  &  &  Atlas \\
    Na-23  &  23.00  &  15.00  & 16.00  &  &  &  &  Atlas \\
    Mg-24  &  20.80  &  41.60  &  9.00  &  &  &  &  Atlas \\
    Al-27  &  21.10  &  12.50  &  6.10  &  29.50  &  6.70  &  8.70 &  RIPL-2 \\
    Si-28  &  20.24  &  58.73  &  5.00  &  &  &  &  Atlas \\
    Ar-40  &  20.90  &  50.00  & 10.00  &  &  &  &  Atlas \\
    Ca-40  &  19.77  &  97.06  &  5.00  &  &  &  &  Atlas \\
    V-51   &  17.93  &  53.30  &  3.62  &  20.95  &  40.70  &  7.15 &  RIPL-2 \\
    Mn-55  &  16.82  &  51.40  &  4.33  &  20.09  &  45.20  &  4.09 &  RIPL-2 \\
    \hline
\end{tabular}
\caption{Giant dipole resonance parameters used with TALYS (as parameters for the Kopecky-Uhl generalized Lorentzian model of the E$1$-strength function): peak energy $E_i$, peak cross section $\sigma_i$ and width $\Gamma_i$ for resonances with a single ($i=0$) or a split peak ($i=0,1$). Default values from the RIPL-2 database are replaced, if available, with the total cross section parameters from the atlas of GDR parameters. Note that for isotopes not listed, as well as for higher order contributions, TALYS uses a compilation of formulas listed in \cite{talys1.6-manual}.}
\label{tab:GDRparameters}
\end{table}

Yet another model is that by Kossov~\cite{kossov2002} used in the Geant4 software. It describes the total cross sections for photodisintegration, as well as at higher energies pion production.
Since branching ratios of individual disintegration channels are not modeled, these are taken from TALYS, when the model is used in CRPropa.

In figure~\ref{fig:sigma} we compare cross sections predicted by the three models we used in this work with available measured data for the total photoabsorption cross section of silicon-28 and for the ejection of an $\alpha$-particle from carbon-12 (with subsequent near-immediate decay of the residual beryllium-8 into two more $\alpha$-particles). For completeness, we also show cross sections computed by the publicly released versions of TALYS using their default settings; in particular, TALYS-1.0 with default settings is the photodisintegration model used in CRPropa~2. It can be seen that TALYS with the parameters used in the original paper most closely reproduces the total cross section data, but PSB and Kossov also give acceptable results, whereas TALYS used with its default settings predicts much broader and lower peaks than observed. On the other hand, all versions of TALYS (especially TALYS-1.0) largely overpredict cross sections for $\alpha$-particle ejection, as does the Kossov model with TALYS branching ratios (though at higher photon energies), whereas the PSB model neglects it altogether. Note that for kinematical reasons the cross sections at the lowest photon energies are the ones most relevant for UHECR propagation.

\section{Models for extragalactic background spectrum\label{sec.EBLmodels}}
The spectrum of the diffuse extragalactic background radiation spans over 20 decades in energy, from radio waves up to the high-energy gamma ray photons. It consists of light emitted at all epochs,
modified by redshifting and dilution due to the expansion of the
universe. The cosmic microwave background (CMB), the relic blackbody radiation
from the Big Bang, is the dominant background field, followed by
ultraviolet/optical and infrared backgrounds (extragalactic background light, EBL). In this work, several models for EBL have been used; these models are included in the simulation codes for propagation with different choices for considering how their spectral energy distribution evolves in redshift. 

The understanding of the spectral energy distribution and redshift evolution of the EBL requires studying the sources responsible for its production. Several different techniques are used for this purpose. Kneiske et al. \cite{Kneiske:2003tx} report the present-day background intensity using detailed information from galaxy surveys about global quantities as the cosmic star formation rate. The work by Dom\'inguez et al. \cite{Dominguez:2010bv} is observationally based on multiwavelength data. Other authors (as for example Stecker et al. \cite{Stecker:2005qs,Stecker:2006eh}) use ``backward evolution'' of the present day galaxy emissivity.  On the contrary, ``forward evolution'', which begins with cosmological initial conditions and follows a forward evolution with time by means of semi-analytical models of galaxy formation, is used in Gilmore et al. \cite{Gilmore:2011ks}. 

\begin{figure}[t]
    \centering
    \includegraphics[width=.495\textwidth]{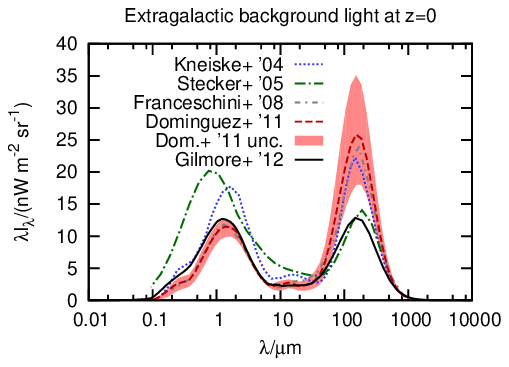}\includegraphics[width=.495\textwidth]{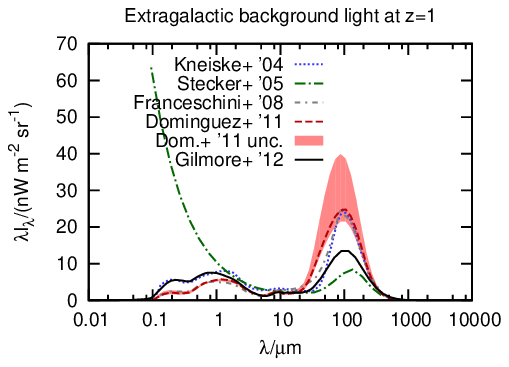}
\caption{Intensity of the EBL at $z=0$ (left) and $z=1$ (right). Wavelengths are given in local physical lengths at the given redshift, and densities are per unit of comoving volume.}
    \label{fig:ebl}
\end{figure}

In figure~\ref{fig:ebl} the intensity of the EBL at $z=0$ (left) and $z=1$ (right) as a function of wavelength is shown, as predicted by the models used in this work (Gilmore 2012 \cite{Gilmore:2011ks} and Dom\'inguez 2011 \cite{Dominguez:2010bv}) and by the default EBL models used in {\it SimProp}~v2r0 and CRPropa~2 (Stecker 2005 \cite{Stecker:2005qs,Stecker:2006eh} and Kneiske 2004 \cite{Kneiske:2003tx} respectively), as well as the Franceschini 2008 model~\cite{Franceschini:2008tp} for comparison.
It can be seen that all recent EBL models are in good agreement concerning the EBL spectrum in the UV and optical region in the local universe, but they still largely differ in the far IR region and at high redshifts. (Note that due to the $1/\epsilon^2$ factor in eq.~\eqref{eq:lambda}, the far IR region is the most relevant to UHECR propagation, and the UV region has very little impact even for the Stecker 2005 model at high redshift where it largely exceeds other models.)

\bibliographystyle{JHEP} 
\bibliography{SimPropCRP}

\end{document}